\documentclass{article}
\usepackage{multicol}

\usepackage{graphicx} 
\usepackage{pdfpages}
\usepackage{caption}

\usepackage{url}

\usepackage[a4paper, margin=2cm, includehead]{geometry} 

\usepackage[all]{nowidow} 
\usepackage[protrusion=true,expansion=true]{microtype} 

\usepackage{natbib}

\begin{document}

\title{\textbf{SBMLtoOdin and Menelmacar: Interactive visualisation of systems biology models for expert and non-expert audiences}}

\author{Leonie J. Lorenz (1), Antoine Andréoletti (1,2,3), Tung V. N. Nguyen (1),\\ Henning Hermjakob (1), Richard G. FitzJohn (4), Rahuman S. Malik Sheriff (1),\\ John A. Lees (1, *)}%

\date{}

\maketitle%
\begin{flushleft}
(1) European Bioinformatics Institute, European Molecular Biology Laboratory, Wellcome Genome Campus, Hinxton, Cambridge, CB10 1SD, United Kingdom, (2) Ecole Polytechnique, Route de Saclay, 91128, Palaiseau Cedex, France, (3) Technical University of Denmark, Anker Engelunds Vej 101, 2800, Kongens Lyngby, Denmark, (4), MRC Centre for Global Infectious Disease Analysis; and the Abdul Latif Jameel Institute for Disease and Emergency Analytics (JIDEA), School of Public Health, Imperial College London, London, W2 1PG, United Kingdom.
\end{flushleft}


\begin{flushleft}
\textbf{Motivation: } Computational models in biology can increase our understanding of biological systems, be used to answer research questions, and make predictions. Accessibility and reusability of computational models is limited and often restricted to experts in programming and mathematics. This is due to the need to implement entire models and solvers from the mathematical notation models are normally presented as.\\
\textbf{Implementation: } Here, we present \texttt{SBMLtoOdin}, an R package that translates differential equation models in SBML format from the BioModels database into executable R code using the R package odin, allowing researchers to easily reuse models. We also present  \texttt{Menelmacar}, a a web-based application that provides interactive visualisations of these models by solving their differential equations in the browser. This platform allows non-experts to simulate and investigate models using an easy-to-use web interface.\\
\textbf{Availability: } \texttt{SBMLtoOdin} is published under open source Apache 2.0 licence at \url{https://github.com/bacpop/SBMLtoOdin} and can be installed as an R package. The code for the \texttt{Menelmacar} website is published under MIT License at \url{https://github.com/bacpop/odinviewer}, and the website can be found at \url{https://biomodels.bacpop.org/}.
\end{flushleft}

\begin{multicols}{2}
\section{Introduction}

Computational models of biological systems can help to increase our mechanistic understanding of biological processes, test hypotheses that would be difficult or expensive to test in a laboratory, and predict the system's behaviour under different conditions. For example, in cancer, computational models of intracellular signalling can uncover changes in molecular interactions, investigate the influence of certain components on others, and predict how a cancer cell would respond to therapy \citep{du_cancer_2015, janes_systems_2024}. Computational models can be implemented in various different coding languages, such as R, Python, or Matlab, but typically require a high level of modeling and computational expertise to be developed, used, and reused. This makes it difficult for researchers who do not have a strong background in programming or mathematics to understand or make use of computational models in biology.

BioModels is a database that hosts about 4000 computational models for biology \citep{glont_biomodels_2018, malik-sheriff_biomodels_2020}. The team from the BioModels database curates models, meaning that they validate that the model submissions by comparing the output to what is described in the corresponding publication. At the moment, roughly 1000 models are curated. The database has a large user community with an average of more than 23,000 unique users per month in 2018 \citep{malik-sheriff_biomodels_2020} and was deemed the most widely used repository for depositing mathematical models in systems biology \citep{stanford_evolution_2015}. Although the BioModels database allows submission in different coding languages, the vast majority of models is deposited in the Systems Biology Mark‐up Language (SBML) \citep{keating_sbml_2020}, which is a machine-readable but non-executable XML file type.

The BioModels database runs in the spirit of the FAIR (findable, accessible, interoperable and reusable) Guiding Principles \citep{wilkinson_fair_2016}. While FAIR principles have been found to have a predominantly positive influence on reuse and quality, lack of skills and access to infrastructure have been found to be a major barrier for data reuse \citep{klebel_academic_2025}. In the case of BioModels, model files are freely accessible on the database. However, accessibility to the model components and dynamics is limited to experts because retrieving information from the SBML file format requires expertise. SBML is a non-executable model exchange format, hence requires the use of a software tool to translate models into executable code. Using these software tools and running models in the resulting coding language requires coding expertise. This work aims to lower these barriers.

\begin{figure*}[ht]
    \centering
    \resizebox{\textwidth}{!}{\includegraphics{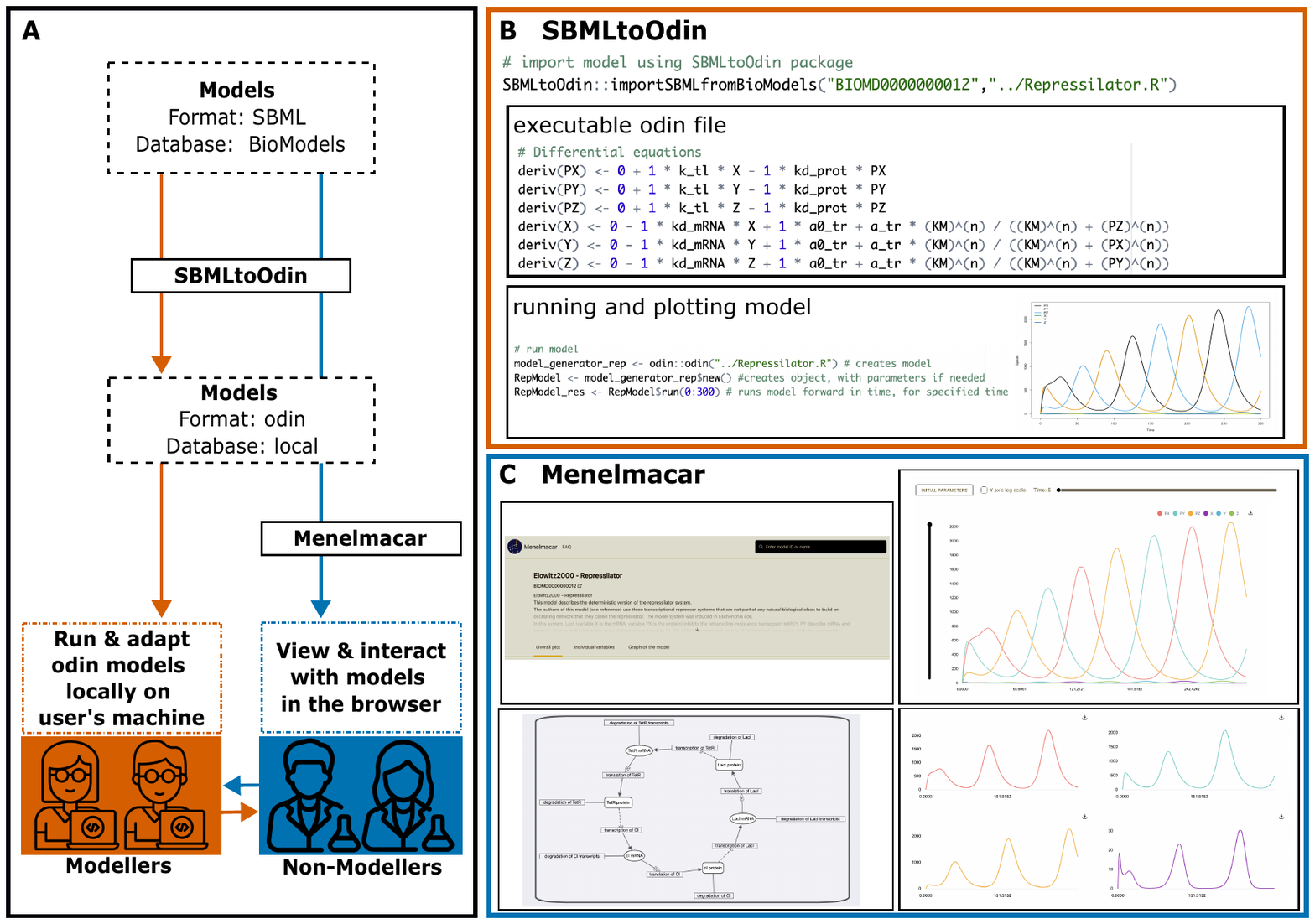}}
    \caption[Pipeline and Website]{\textbf{(A)} \texttt{SBMLtoOdin} translates models in SBML format, typically taken from the BioModels database, into executable odin code. Modellers can then run and adapt odin models locally on their machine (B). For non-modellers, the \texttt{Menelmacar} website provides interactive visualisations of the models (C). \textbf{(B)} Users can translate models from SBML to odin code with just one simple command, the models can then be run, plotted and adapted using the odin R package. \textbf{(C)} Our website \texttt{Menelmacar} (\url{biomodels.bacpop.org}) allows users to interact with the models without needing expertise in coding or modelling.}
    \label{fig:PipelineWebsite}
\end{figure*}

To facilitate access and reuse of computational models in biology, we developed the R Package \texttt{SBMLtoOdin}, which translates ordinary differential equation (ODE) models into an executable format. \texttt{SBMLtoOdin} translates models from SBML format into executable code of the domain-specific language odin \citep{fitzjohn_reproducible_2020}, which can be flexibly called and plotted from R. Modellers can then easily adapt, simulate, and fit odin models in their preferred coding environment, for example R Studio. We ran \texttt{SBMLtoOdin} on all ODE-based curated models of the BioModels database and made interactive visualisations available on our website \texttt{Menelmacar} by using the javascript solver available in odin. Experts and non-experts can interact with models in their browser, using a search bar to find models of interest and using arrows to change parameter values, the impact of which can be observed in the plots. There is no need for any coding expertise for interacting with the models in the browser. A flowchart of our pipeline can be found in Figure \ref{fig:PipelineWebsite}A.%

\section{Implementation}

\subsection{Translating models from SBML to odin}

Our R package \texttt{SBMLtoOdin} takes computational models in SBML format and translates them into executable models in the domain-specific programming language odin \citep{fitzjohn_reproducible_2020}. 

The SBML file format is an XML format that is commonly used in the systems biology community \citep{keating_sbml_2020}. The format of SBML makes it easy to share model definitions with other researchers because the files are small, and the format is independent of the users' operating systems or coding languages. However, SBML is not directly executable itself, meaning that the format has to be translated into a coding language before users can run the models. This process consists of extracting equations, variables, and parameters from the SBML file using an XML library, restructuring the information for the coding language of choice, defining a solver for running the model, and finally visualizing the results. Our R package \texttt{SBMLtoOdin} automates the extraction of the models and their implementation in a coding language with a readily available solver, meaning that the users do not have to do any of these steps manually. \texttt{SBMLtoOdin} uses libSBML \citep{bornstein_libsbml_2008},  a library for importing and exporting models in SBML format, to import the SBML files into R, extracts the necessary model information from the R objects, and translates equations and parameters into executable code.

We used the R package odin to create executable implementations of the models. Odin is a domain-specific programming language for state space models \citep{fitzjohn_reproducible_2020}. Odin code looks similar to R code but gets transpiled to C++, leveraging the speed of this coding language, but maintaining the readability and formatting of the underlying equations. 

It requires just one simple line of code to translate a model using \texttt{SBMLtoOdin} (Figure \ref{fig:PipelineWebsite}B). There are two functions for translating models: \texttt{importSBMLfromBioModels} and \texttt{importSBMLfromFile}. \texttt{importSBMLfromBioModels} is a function for importing models from the BioModels database using the BioModels ID, which downloads the corresponding SBML file using the BioModels' API. \texttt{importSBMLfromFile} translates models that are saved on the user's computer, which can be downloaded models from the BioModels database or models from other origin. Both functions then translate the SBML file into odin code. Users can then simulate the models using the R package odin, adapt the model code, or fit the models to data using the R package mcstate \citep{fitzjohn_reproducible_2020}. An example of how to import a model from BioModels using the \texttt{importSBMLfromBioModels} function, including a snippet of the produced odin code and a plot can be found in Figure \ref{fig:PipelineWebsite}B.

We tested \texttt{SBMLtoOdin} on the SBML test suite \citep{hucka_sbmlteam_2022}, which is a collection of SBML models along with expected simulation results. \texttt{SBMLtoOdin} currently does not support compartment sizes unequal to one, algebraic rules, conversion factors, assignment rules for species (unless these species are boundary conditions), events for species and events with self-referencing, and fast reactions (which have been removed from SBML in version 3.2). We tested \texttt{SBMLtoOdin} on the first 1,000 semantic cases of the SBML test suite, excluding those with above-mentioned features from the analysis. \texttt{SBMLtoOdin} translated all models successfully and the solver produced the correct results within an error margin of $10^{-5}$. 

\subsection{Visualizing models in the browser}

Our web-based application \texttt{Menelmacar} (\url{biomodels.bacpop.org}) (Making Execution of (Nearly) Every Life-science Model ACcessible to All Researchers) can be used for visualizing and interacting with computational models from the BioModels database. It is a lightweight front-end browser application written using the vue framework, and runs without a backend or server, making its future maintenance simple.

To generate the content of the website, we ran \texttt{SBMLtoOdin} on the curated models from BioModels. To avoid errors in translating models, we filtered the 1,096 curated models by the above-mentioned features and translated the models using \texttt{SBMLtoOdin}, which resulted in 590 models. Then we transpiled the odin code to javascript, which allows us to run and display the models in the browser, with the model solver being run directly by the user's browser. We also translated the SBML model files to JSON format using SyBLaRS \citep{balci_syblars_2022}, which allows us to display the model structure as a graph. In this graph, nodes represent variables of the model and edges represent interactions between variables. 

The landing page of the website contains a search bar, which supports search using the BioModels ID. This page also links to example models and has an FAQ section. 


Models can be viewed in three different ways: as a time trajectory of all model variables in one plot; a time trajectory separately for each model variable; a graph representing the network structure and connections of variables of the model. All of these visualisations are interactive. Users can, for example, change start parameters and the axes dimensions and observe changes in the model behavior. The plots and the model graph can also be downloaded. An example of the model trajectories on the website is shown in Figure \ref{fig:PipelineWebsite}C.

The BioModels database created links from their website to \texttt{Menelmacar} for all the models that we currently support (e.g. \url{https://www.ebi.ac.uk/biomodels/BIOMD0000000012}).

User feedback is important in ongoing development and validation of computational models, and engaging more users in this process will improve the reliability of model reuse. To enable this we have added a link to our github repository on the bottom of each model page. Through this link, users can report errors in models using a simple form that already quotes the model ID.

\section{Discussion}

A recent review hast found that FAIR principles have been adopted by many databases and researchers and the overall impact on citations and reuse appears to be positive \citep{klebel_academic_2025}. However, the authors also find that accessibility is not everything: lack of skill is a major hurdle in the reuse of open/FAIR data, negatively impacting equity and inclusion. This emphasizes the need for lowering the barriers for accessibility and reusability, beyond the definitions of FAIR principles. In our view, this is most likely even more pressing for mathematical modelling in biology, which requires a very specific set of skills and therefore has an intrinsically higher barrier. 

While the uptake of the BioModels database in the systems biology community is successful \citep{malik-sheriff_biomodels_2020, stanford_evolution_2015}, model sharing is much less common in other mathematical modelling communities in biology, such as epidemiological modelling. During the Covid-19 pandemic for example, approaches were developed to facilitate model sharing and interoperability through the BioModels database, specifically for epidemiological models \citep{ramachandran_fair_2022}. However, countless research groups around the globe made efforts to develop mathematical models for predicting the course of the pandemic, often using very similar methods but with little communication even within countries \citep{streicher_need_2025}. Model sharing and reuse could not only reduce workload in future pandemics but also potentially lead to faster response times and more accurate predictions, which are then easier for external groups to validate.

With our tools, we hope to contribute to an increase in model sharing and reuse. By bridging a model file type that is predominantly used in the systems biology community (SBML) and a coding language that is predominantly used by epidemiological modellers (odin), we have lowered barriers to model sharing between the communities and between epidemiological modellers. A possible future extension that might further encourage epidemiological modellers to use the BioModels database is the implementation of a tool that translates odin files into SBML format.

Other future work could be to extend the pipeline to support non-ODE models, specifically stochastic models. Stochastic models account for randomness and uncertainty, and they can be used as an alternative or in addition to deterministic (e.g. ODE) models. This would be an important extension because stochastic models are common in epidemiological modelling and it would be feasible since stochastic models are supported by the odin packages \citep{fitzjohn_reproducible_2020}.

Overall our set of packages facilitates access and reuse of mathematical models in biology. Our R package \texttt{SBMLtoOdin} automatizes the translation of SBML models into odin models, which can then easily be run and plotted. Our website  \texttt{Menelmacar} provides easy access to models for modellers and non-modellers alike by providing interactive visualisations of the models.

\section{Competing interests}
No competing interest is declared.

\section*{Funding}
This work was supported by the European Molecular Biology Laboratory, European Bioinformatics Institute [to L.J.L., R.S.M.S., and J.A.L.] and the French Embassy in London [to A.A.].

\section*{Data availability}
Code for \texttt{SBMLtoOdin} is available on GitHub (\url{https://github.com/bacpop/SBMLtoOdin}). Code for Menelmacar website is also available on GitHub (\url{https://github.com/bacpop/odinviewer}). 



\section{Acknowledgments}
We thank Andrea Epifani and Zeqing Lu from Gomoku Studio for improving the design of the  \texttt{Menelmacar} website. We thank Joel Hellewell for his advice on developing R packages and Alireza Tajmirriahi for some last-minute support for the website.

\bibliographystyle{abbrvnat}

\bibliography{references}

\end{multicols}
\end{document}